\documentstyle[nato,numreferences,epsf,graphicx]{crckapb}
\newcommand{\tr}{{\rm Tr}}
\newcommand{\be}{\begin{equation}}
\newcommand{\ee}{\end{equation}}
\newcommand{\bea}{\begin{eqnarray}}
\newcommand{\eea}{\end{eqnarray}}

\begin{opening}
\title{$SO(3)$ versus $SU(2)$ lattice gauge theory}

\author{Philippe de Forcrand}
\institute{Inst. f\"ur Theor.\ Physik,
       ETH, CH-8093 Z\"urich, Switzerland \\
and Theory Division, CERN, CH-1211 Geneva 23, Switzerland }

\author{Oliver Jahn}
\institute{Inst. f\"ur Theor.\ Physik,
       ETH, CH-8093 Z\"urich, Switzerland }

\end{opening}

\begin{document}

\begin{abstract}
We consider the $SO(3)$ lattice gauge theory at weak coupling,
in the Villain action. We exhibit an analytic path in coupling space 
showing the equivalence of the $SO(3)$ theory with $SU(2)$ summed
over all twist sectors. This clarifies the ``mysterious phase'' of
$SO(3)$. As order parameter, we consider the dual string tension
or center vortex free energy,
which we measure in $SO(3)$ using multicanonical Monte Carlo.
This allows us to set the scale, indicating that ${\cal O}(700)^4$
lattices are necessary to probe the confined phase.
We consider the relevance of our findings for confinement in 
other gauge groups with trivial center.
\end{abstract}

\section{Motivation}

Our main motivation for a numerical study of the $SO(3)$ gauge theory on the
lattice is to clarify an apparent paradox.

On one hand, $SO(3) = SU(2)/Z_2$, so that the $SO(3)$ theory is an $SU(2)$
theory in the adjoint representation. On the lattice, the Yang-Mills action
$-\frac{1}{4}\int d^4x F_{\mu\nu}^2$ can be obtained as the naive continuum
limit of a plaquette action taken in any representation. In the usual Wilson
action $\beta_F \sum_P (1 - \frac{1}{2}\tr_F U_P)$, the trace of the plaquette
matrix $U_P$ is taken in the fundamental representation. The $SO(3)$ theory
corresponds to action $\beta_A \sum_P (1 - \frac{1}{3}\tr_A U_P)$, with the
trace taken in the adjoint representation. The universality of the continuum
fixed point leads us to believe that the $SO(3)$ and $SU(2)$ theories are 
equivalent not just in the naive continuum limit, but also non-perturbatively.
Thus, $SO(3)$ should confine (at low temperature $T < T_c$), just like $SU(2)$.

On the other hand, $Center(SO(3)) = {\bf 1}$. There is no center symmetry to
break in the case of $SO(3)$. This means that the well-accepted connection 
between center symmetry breaking and the deconfinement transition \cite{SY}
cannot apply. Another order parameter must be found. In the process,
we may learn some general lesson about the gauge group structure necessary
to sustain confinement.

Our attention will focus on topological excitations common to both $SU(2)$ 
and $SO(3)$: center vortices. Remarkably, a group with a trivial center may
still support center vortices. A center vortex is a two-dimensional topological
excitation. To prevent its action from diverging, it is a pure gauge at
$r=\infty$, characterized by the gauge transformation $g(\theta)$ applied to the
trivial vacuum. In $SU(N)$, the non-trivial topology comes from the possibility
that $g(\theta+2\pi) = g(\theta) \exp(i \frac{2\pi}{N} k), k=0,..,N-1.$
The integer $k$ is the ``twist'', defined mod $N$, and the Wilson loop at
$r=\infty$ takes value $W(r=\infty) = g^{-1}(0) g(2\pi) = \exp(i \frac{2\pi}{N} k)$.
Notice that both $W(r=\infty)$ and the gauge field 
$A_\mu(r=\infty,\theta) = g^{-1}(\theta)\partial_\mu g(\theta)$
are unchanged if $g(\theta)$ is multiplied by a $Z_N$ center element.
This implies that our topological excitations are characterized by equivalence
classes $SU(N)/Z_N$ of gauge transformations. Twist arises from the non-trivial
mapping of the $r=\infty$ circle to these equivalence classes:
$\Pi_1(SU(N)/Z_N) = Z_N$. For a general non-Abelian gauge group $G$, center vortices
exist if $\Pi_1(G/Center(G)) \neq {\bf 1}$. Thus, both $SU(2)$ and $SO(3)$ 
admit center vortices. 

One may also realize this fact by recalling that center vortices arise as 
local gauge singularities after center gauge fixing \cite{Pepe}, where center
gauge is just the Landau gauge in the adjoint representation: the gauge condition
only makes use of the $SO(3)$ content of the gauge links, so that the gauge
singularities only depend on the $SO(3)$ part.

\section{Center vortices in $SU(2)$}

The relevance of center vortices for confinement in $SU(2)$ has been studied 
numerically in two ways. One can fix the gauge to a maximal or Laplacian center
gauge, where center vortices can be identified as gauge singularities \cite{Pepe}
or as P-vortices \cite{Green} and center projection can be performed.
One can also study center vortices in a gauge-invariant way via the 't Hooft
loop \cite{tH}, since a 't Hooft loop of contour ${\cal C}$ creates a fluctuating
center vortex sheet bounded by ${\cal C}$ \cite{LAT01}. In the latter case,
a 't Hooft loop of maximal size $L_\rho \times L_\sigma$ is equivalent to enforcing
twisted boundary conditions \cite{tH} in the orthogonal $\mu \nu$ plane. A simple way to
see this is to observe that the twist matrices $\Omega_\mu, \Omega_\nu$ satisfy
(for $SU(2)$) $\Omega_\mu \Omega_\nu = - \Omega_\nu \Omega_\mu$. Therefore,
a Wilson loop of size $L_\mu \times L_\nu$ picks up a factor 
$\Omega_\mu \Omega_\nu \Omega_\mu^\dagger \Omega_\nu^\dagger = -{\bf 1}$.
In other words, twisted b.c. multiply the largest Wilson loop at ``$r=\infty$''
by a center element, precisely as a center vortex does: twisted b.c. ($tbc$)
create one center vortex (mod $2$) as compared to periodic b.c. ($pbc$).

\begin{figure}
\centering
\mbox{\hspace{-1cm}\epsfxsize=10cm\epsfysize=6.8cm\epsffile{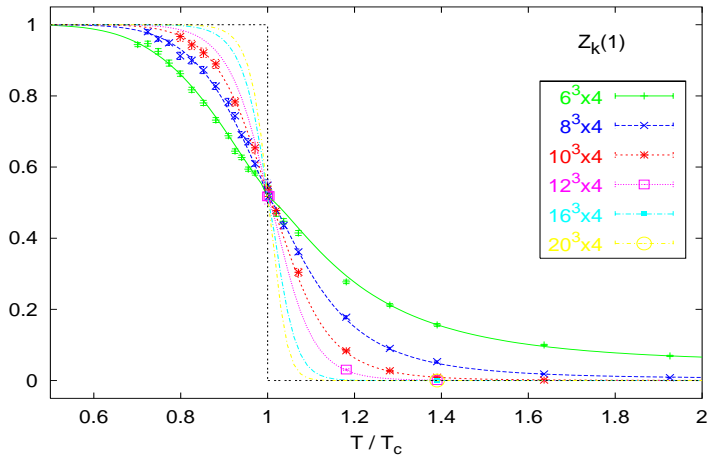}}\vspace{-5mm}
\caption{The effect of temporal twist in $SU(2)$: ratio $Z_{tbc}/Z_{pbc}$ 
as a function of temperature, for various lattice sizes (from \protect\cite{LvS}).}
\label{ZkTc}
\end{figure}

The ratio of partition functions $Z_{tbc}/Z_{pbc} = \exp(-F_{CV})$ probes the
free energy $F_{CV}$ of a center vortex. Note that UV divergences cancel out 
in the ratio, so that it is a well-defined continuum quantity. On a hypercubic
$L^4$ lattice, this ratio rapidly tends to $1$ as $L$ increases above 
$\sim 0.8$ fm \cite{KT}. At finite temperature, considering electric twist
(i.e. in a temporal plane), this ratio is an order parameter for confinement 
\cite{LvS}. Below $T_c$, the correlation length $\xi$ is finite and boundary
conditions do not matter if the system size $L$ is $\gg \xi$: 
$Z_{tbc}/Z_{pbc} \rightarrow 1$. Above $T_c$, center vortices are ``squeezed''
by the compact Euclidean time. Their free energy increases as 
$\tilde\sigma(T) L^2$, generating an area law for the spatial 't Hooft loop.
The prefactor $\tilde\sigma(T)$ is the dual string tension, which varies with
$T$ and vanishes as $T \rightarrow T_c^+$ with Ising-like critical exponents
\cite{PRL}. Since we claim to measure in $Z_{tbc}/Z_{pbc}$ a physical quantity,
it should be measurable as well in the $SO(3)$ theory. We will see that this
is indeed the case. $Z_{tbc}/Z_{pbc}$ is an order parameter also for $SO(3)$,
and $\tilde\sigma(T)$ provides a way to measure the temperature also in 
that theory.

\section{$SU(2)-SO(3)$ phase diagram and the $SO(3)$ ``mysterious phase''}

Contrary to the Wilson action $\beta_F \sum_P (1 - \frac{1}{2}\tr_F U_P)$
which enjoys a smooth crossover from strong to weak coupling, the adjoint
action $\beta_A \sum_P (1 - \frac{1}{3}\tr_A U_P)$ produces a bulk, first-order
phase transition at $\beta_A \approx 2.5$, which prevents an arbitrary 
decrease of $\beta_A$ while studying the weak coupling phase relevant for
the continuum limit. The common wisdom is therefore that $SO(3)$ gives the same
continuum physics as $SU(2)$, but that the lattice spacing is kept small by
the necessity of staying on the weak coupling side of the bulk transition.
One purpose of our study is to check this scenario. There is room for skepticism,
because the phase diagram of a mixed $(\beta_F,\beta_A)$ action, first studied
by Bhanot and Creutz \cite{BC}, shows the weak coupling $\beta_F=0$ (i.e. $SO(3)$) 
phase qualitatively separated from the $SU(2)$ theory by a line of first-order 
phase transitions: there is no analytic $(\beta_F,\beta_A)$ path connecting
the two. The possibility that the two theories are different has actually been
proposed, with numerical results to support it, in \cite{LR}.

Another feature of $SO(3)$ encouraging such speculation is the existence of 
a ``mysterious phase'', reported by Datta and Gavai \cite{DG}. This phase appears
in all respects similar to the ``normal'' phase, except for the Polyakov loop $P$:
instead of $\frac{1}{2} \tr_F P$ approaching $\pm 1$ as $\beta_A \rightarrow \infty$,
it approaches zero (so that, in the adjoint representation, $\tr_A P \rightarrow -1$
instead of $+3$). Tunneling between the two phases appears extremely infrequent,
increasingly so at higher $\beta_A$. One wonders how the $SO(3)$ theory could 
have two distinct continuum limits, both similar to $SU(2)$.

To investigate this puzzle, we consider the Villain action for the $SO(3)$ theory,
which associates a discrete variable $\alpha_P = \pm 1$ to each plaquette and
has action $S = -\beta_V \sum_P \alpha_P \frac{1}{2}\tr_F U_P$. Note that $\alpha_P$  
can be integrated analytically, giving 
$S = -\sum_P \log \rm{cosh}(\frac{\beta_V}{2} \tr_F U_P)$,
which is manifestly invariant under a change of sign of $\tr_F U_P$, like the
usual adjoint action $\beta_A \sum_P (1 - \frac{1}{3} \tr_A U_P) = 
\frac{4}{3}\beta_A \sum_P (1 - (\frac{1}{2}\tr_F U_P)^2)$.
Therefore, we deal with a genuine $SO(3)$ action, which shows the same bulk
first-order transition (at $\beta_V \approx 4.47$) and the same ``mysterious phase''
\cite{DG}. The advantage of the Villain choice is that the connection 
$SO(3) \leftrightarrow SU(2)$ is easier to display.

\section{Solving the puzzle: an analytic path between $SU(2)$ and $SO(3)$}

This section contains no new results. We are simply pulling together observations
previously made by various authors. Even the analytic path between $SO(3)$ and
$SU(2)$ was already hinted at by Halliday and Schwimmer \cite{HS}.

The first observation, originally due to Mack and Petkova \cite{MP}, was made
precise by Tomboulis and Kovacs \cite{KT2}. The $SU(2)$
partition function for the Wilson action can be exactly rewritten as an $SO(3)$
partition function in Villain form, with the additional constraint that 
$\alpha_P$-monopoles are forbidden. Namely:
\be
Z_{SU(2)} \equiv \int {\cal D}U e^{\beta \sum_P \tr_F U_P} \\
          = c \!\!\sum_{\alpha_P = \pm 1} \int {\cal D}U e^{\beta \sum_P \alpha_P \tr_F U_P} \!\!
\prod_{\rm{cubes}} \! \delta(\prod_6 \alpha_P - 1)
\label{KTequiv}
\ee
where $c$ comes from the normalization of the $\alpha_P$ integration.

The second observation is due to Alexandru and Haymaker \cite{AH}, and is
a refinement upon the first. While eq.(\ref{KTequiv}) applies in an infinite
volume, on a 4-torus one needs some additional global constraints on the $\alpha_P$
variables:
\be
N_{\mu\nu} \equiv \prod_{P~\in~\rm{plane}~\mu\nu} \alpha_P = +1
\label{AHtwist}
\ee
for each product of $L_\mu \times L_\nu$ variables $\alpha_P$ in each $\mu\nu$ plane.
Only configurations satisfying (\ref{AHtwist}) can be mapped to $SU(2)$ 
configurations with periodic b.c. Note that, in the absence of $\alpha_P$-monopoles,
$\prod_{P~\in~\rm{plane}~\mu\nu} \alpha_P$ is the same for every parallel plane
$\mu\nu$. Therefore, eq.(\ref{AHtwist}) defines 6 constraints only.
If one imposes $N_{\mu\nu} = -1$ for some $\mu\nu$ orientation, then the 
mapping (\ref{KTequiv}) is preserved, except that the $SU(2)$ b.c. are
{\em twisted} in the $\mu\nu$ plane.

It is then straightforward to remove the constraints (\ref{AHtwist}) by 
summing over all $2^6$ $N_{\mu\nu}$ possibilities. One obtains:
\be
\sum_{\alpha_P = \pm 1} \int {\cal D}U e^{\beta \sum_P \alpha_P \tr_F U_P}
\prod_{\rm{cubes}} \delta(\prod_6 \alpha_P - 1) = 
c^{-1} \!\!\! \sum_{\rm{all~twist~sectors}} \!\!\! Z_{SU(2)}
\label{twistequiv}
\ee
The $SO(3)$ Villain partition function, with monopoles removed, is just
the sum of $SU(2)$ partition functions over all possible twisted b.c.

The ``mysterious phase'' then must be one (or more) of these twisted $SU(2)$
sectors. Indeed, one knows how to construct a so-called ``twist-eater'',
a groundstate with zero action and twisted b.c. \cite{twisteater}.
Suppose $N_{xt}=-1$ is enforced through $\alpha_P=-1$ for all $xt$ plaquettes
at $(1,y,z,1)$, $+1$ otherwise. Then, set all links to ${\bf 1}$, except
$U_x(1,y,z,t)= i \sigma_1$ and $U_t(x,y,z,1) = i \sigma_2$.
This choice yields $\frac{1}{2} \tr_F U_P = +1$, except for $x=t=1$ in 
$xt$ planes, where $U_P=-{\bf 1}$. This minus sign cancels with that of
$\alpha_P$ to give zero action. Note that the two Polyakov loops $P_x$ and
$P_t$ in the $x$ and $t$ directions are now $i \sigma_1$ and $i \sigma_2$,
satisfying $\tr_F P = 0$.

\begin{figure}
\centering
\mbox{\epsfxsize=10cm\epsfysize=6.5cm\epsffile{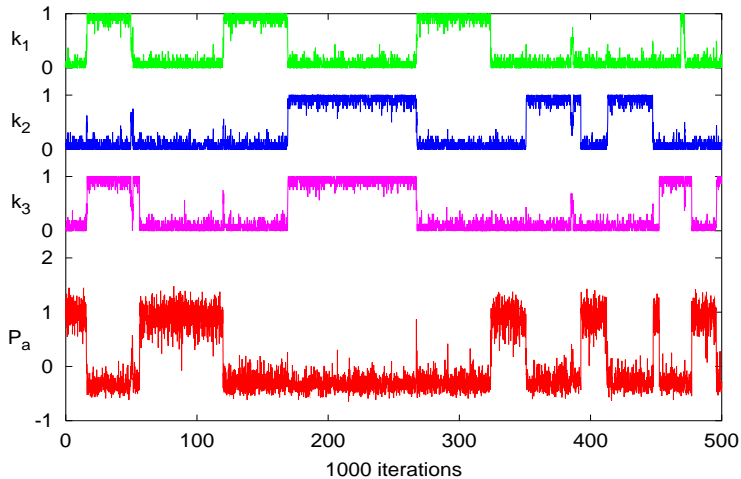}}
\caption{Monte Carlo history of the adjoint Polyakov loop (bottom) and
of the 3 electric twist variables (top).
The trace of the adjoint Polyakov loop is negative whenever twist is present
($4^4$ lattice, $\beta=4.5$).}
\label{MChist}
\end{figure}

Therefore, we guess that, whenever the $SO(3)$ system is in the mysterious
phase characterized by $\tr_F P_t \sim 0$ (i.e. $\tr_A P_t \sim -1$), twist
in the $\alpha_P$ variables must be present in some of the temporal planes.
This proves to be correct, as illustrated in Fig.~\ref{MChist}, where we show
the Monte Carlo history of $\tr_A P_t$, accompanied by that of $k_{x,y,z}$,
where 
\be
k_x \equiv \frac{1}{2} \biggl(1 - \frac{1}{L_y L_z} \sum^{L_y L_z} N_{xt} \biggr)
\label{twistdef}
\ee
is the average, mapped to the interval $[0,1]$, of the $\alpha_P$-twist $N_{xt}$ in each
$xt$ plane. This averaging is necessary because our $SO(3)$ action does not
forbid $\alpha_P$-monopoles. They have a small ($\propto e^{-m \beta}$) density, 
which allows $N_{xt} = \prod_P \alpha_P$ to fluctuate from one parallel plane to another,
so that $k_{x,y,z}$ fluctuate near $0$ or near $1$.
Note also the very slow Monte Carlo dynamics, which prevent an ergodic sampling
of configuration space on all but the smallest lattice sizes 
unless one resorts to special strategies, see sec.\ \ref{results}.

By measuring Polyakov loops in other directions, one can resolve whether 
one or more of the $k_{x,y,z}$ are near 1, and obtain a one-to-one mapping 
between Polyakov loops and twist sectors \cite{tocome}.

\begin{figure}
\centering
\epsfxsize=12cm\epsfysize=6cm\epsffile{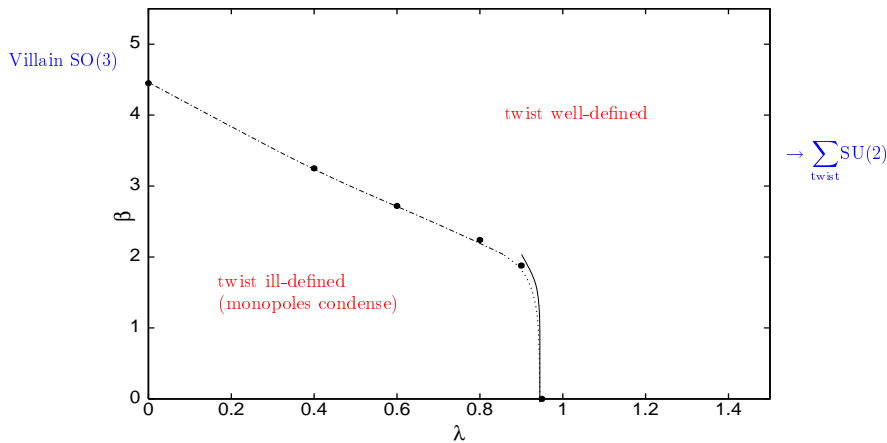}\vspace{-3mm}
\caption{Phase diagram in the coupling plane $(\lambda,\beta)$, where $\lambda$
is the monopole chemical potential and $\beta$ the $SO(3)$ gauge coupling
(adapted from Ref.\protect\cite{DG}).}
\label{DG5}
\end{figure}

Completing an analytic path between $SO(3)$ and $SU(2)$ requires one last,
simple step. We need to allow again $\alpha_P$-monopoles, which can be done
via a chemical potential $\lambda$, giving the partition function
\be
Z = \sum_{\alpha_P = \pm 1} \int {\cal D}U \exp\Bigl(\beta \sum_P \alpha_P \tr_F U_P
+ \lambda \sum_{\rm{cubes}} \prod_6 \alpha_P \Bigr)
\ee
In the limit $\lambda \rightarrow +\infty$, $\alpha_P$-monopoles are forbidden
and eq.(\ref{twistequiv}) holds. The $SO(3)$ Villain action which we study 
corresponds to $\lambda=0$. 
The phase diagram in the $(\lambda,\beta)$ coupling plane
has been studied in \cite{DG}. We reproduce it in Fig.~\ref{DG5}.
At strong coupling, monopoles condense: their density is ${\cal O}(1)$,
which introduces strong fluctuations in $\prod_P \alpha_P$ from one parallel
plane to another. Variables $k_{x,y,z}$ fluctuate around $\frac12$, and the
concept of twist is ill-defined. On the other hand, at weak coupling,
monopoles are exponentially suppressed, long-range order appears among the
$\alpha_P$ variables, and twist sectors become well-defined. There is no
non-analyticity between $(\lambda=0,\beta > \beta_c \approx 4.47)$, which is 
the weak-coupling
$SO(3)$ Villain theory under study, and $\lambda=+\infty$, which is equivalent
to $SU(2)$ summed over all twist sectors as per eq.(\ref{twistequiv}). Therefore,
the non-perturbative continuum physics of $SO(3)$ is the same as that of $SU(2)$,
summed over all twist sectors.

What remains to do is to put this statement on a quantitative footing, by 
matching the lattice spacings $a(\beta)$ in $SO(3)$ and $SU(2)$. We accomplish
this next by measuring the dual string tension $\tilde\sigma(T)$ in both theories.

\section{The order parameter and its measurement}
\label{results}

Defining an order parameter for confinement in $SO(3)$ is not a trivial matter,
as emphasized by Smilga \cite{Smilga}. The trace of the $SU(2)$ Polyakov loop,
like that of all Wilson loops, is identically zero in the fundamental representation.
In the adjoint representation, the trace of the Polyakov loop is always non-zero,
since adjoint charges are screened by gluons at any temperature.
Here, we use the dual string tension $\tilde\sigma(T)$, which is zero in the
confining phase $T < T_c$, positive and increasing with $T$ in the deconfined
phase $T > T_c$, and which we already showed to be a good order parameter
in the $SU(2)$ theory (see \cite{PRL,LvS} and Fig.~\ref{ZkTc}).

In $SU(2)$, on a lattice of physical size $L^3 \times T^{-1}$, the
dual string tension can be obtained from the ratio of twisted (in a temporal
$L \times T^{-1}$ plane) and periodic partition functions:
\be
\tilde\sigma(T) = \lim_{L\rightarrow \infty} -(1/L^2) \log Z_{tbc}/Z_{pbc}
\label{extrapol}
\ee
The presence or absence of twist is imposed by hand, by a specific choice of
boundary conditions. In $SO(3)$, twist sectors are {\em automatically} summed
over, and twist becomes an observable, similar to the usual topological charge
in $SU(2)$. Sectors of different twist, which would be disjoint in the 
continuum limit, are connected at finite $\beta$ via saddle points which are
lattice artifacts, analogous to topological ``dislocations''. The assignment
of an integer twist (in each plane) to a given configuration is subject to
ambiguities similar to the assignment of a topological charge. Fortunately,
we work at a very weak coupling, so that such ambiguities have a negligible
impact in our case: the twist variables $k_{x,y,z}$ (eq.(\ref{twistdef}))
stay near $0$ or $1$. The price we have to pay is that the twist sectors
are separated by very high action barriers $\Delta S$, nearly impassable
for an ordinary Monte Carlo algorithm.

\begin{figure}
\centering
\mbox{\hspace{5mm}\epsfxsize=10cm\epsfysize=7.9cm\epsffile{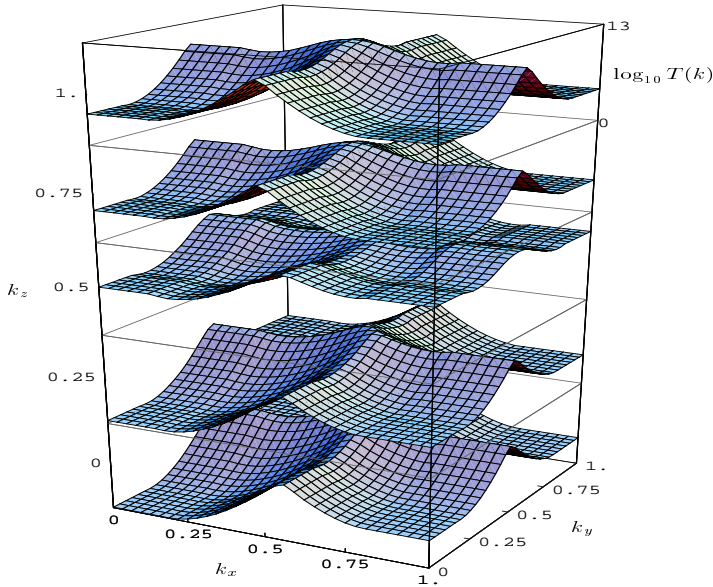}}\vspace{-2mm}
\caption{The three-dimensional reweighting table used in our multicanonical
Monte Carlo. No twist is at bottom left, twist in all 3 planes at top right.
The table enhances the probability of sampling the saddles between twist
sectors. Its entries vary by 13 orders of magnitude 
($8^3\times 4$ lattice, $\beta=4.5$).}
\label{3dtable}
\end{figure}

Our strategy to solve this technical problem is multicanonical sampling \cite{BN}.
Instead of assigning probability $\frac{1}{Z}e^{-S}$ to each configuration,
we multiply this probability by a factor $T(k)$ which favors the saddle points
between twist sectors $k\sim 0$ and $k\sim 1$, tuned such that the resulting
sampling probability is flat from one sector to the next. The Monte Carlo
process then amounts to a free diffusion across twist sectors, with
dynamics accelerated exponentially (by a factor $\sim e^{\beta \Delta S}$).

\begin{figure}
\centering
\mbox{\epsfxsize=9cm\epsfysize=5.9cm\epsffile{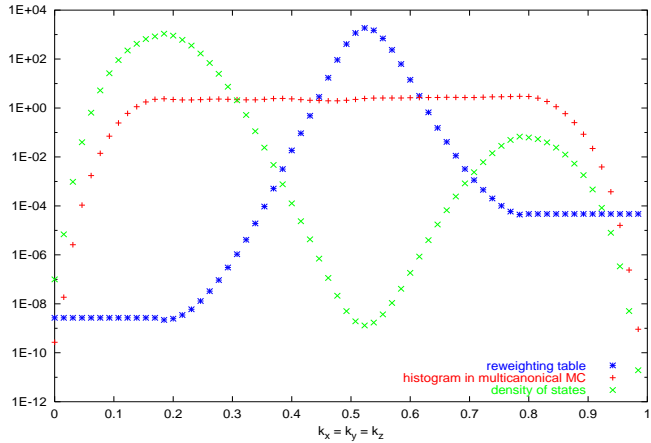}}
\caption{Cut of the three-dimensional reweighting table along its diagonal (blue *).
The result of multicanonical Monte Carlo sampling is a nearly flat histogram
(red +). \protect\nolinebreak The density of states shows the dominant twist-0 and the smaller
twist-3 sectors (green x).}
\label{diag}
\end{figure}

In our case, we want to facilitate the Monte Carlo sampling of twist in
all 3 temporal planes. This leads us to a multicanonical sampling with
a 3-dimensional reweighting factor $T(k_x,k_y,k_z)$ which can be represented 
by a table since
$k_{x,y,z} = \hat{k}/L^2$, $\hat{k} = [0,..,L^2]$.
This table is constructed iteratively, using converged values on small
lattices to form starting values on larger ones. Such a table is displayed
in Fig.~\ref{3dtable}, for an $8^3 \times 4$ lattice at $\beta=4.5$.
Although $\beta$ has been chosen as small as possible (the bulk transition
to the strongly coupled phase occurs at $\beta_c \approx 4.47$), the
necessary enhancement of the saddle points reaches $10^{13}$, as seen more
clearly in Fig.~\ref{diag}, which is a $1d$-cut of the same table along its
diagonal (from $k_{x,y,z}=(0,0,0)$ to $(1,1,1)$). There, the resulting
flatness of the Monte Carlo sampling probability is clearly visible.
The measured density of states shows two well-separated peaks, corresponding
to twist sectors $(0,0,0)$ (analogous to $SU(2)~pbc$) and $(1,1,1)$
(analogous to $SU(2)~tbc$ in all 3 temporal planes).
The strong suppression of the saddle point confirms that ordinary
Monte Carlo sampling would remain hopelessly ``stuck'' in one sector,
as observed in earlier studies. However, in spite of the great
multicanonical acceleration, the Monte Carlo evolution of the twist
variables $k_{x,y,z}$ is still slow, and simulating a $10^3 \times 4$
lattice remains beyond the edge of our computer resources.


This prevents us from reaching spatial sizes large enough for a
reliable extrapolation of eq.(\ref{extrapol}) to $L \rightarrow \infty$.
Our extrapolation depends on the ansatz we choose for the finite-size
effects. Nevertheless, we can still calibrate the $SO(3)$ lattice spacing
$a(\beta_{SO(3)}=4.5)$, by comparing our $L=4,6,8$ results with those
obtained in $SU(2)$ on the same lattice sizes $L^3 \times 4$.
In both theories, our observables are $Z_{tbc}^k/Z_{pbc} \equiv \exp(-F_{CV}^k)$,
where $k=1,2,3$ refers to twisted b.c. in 1, 2 or 3 temporal planes.
In $SO(3)$, these observables are obtained from the Monte Carlo distribution
of the twist variables $k_{x,y,z}$. This distribution falls into $2^3$
distinct, well-separated peaks, as in Fig.~\ref{diag}, and we 
measure the ratio of the population around each peak to that around the
$pbc$ peak at $k_{x,y,z} \sim 0$. In $SU(2)$, these ratios are obtained
from different Monte Carlo simulations, following the method of \cite{PRL}.

\begin{figure}
\centering
\mbox{\hspace{5mm}\epsfxsize=11cm\epsfysize=6.1cm\epsffile{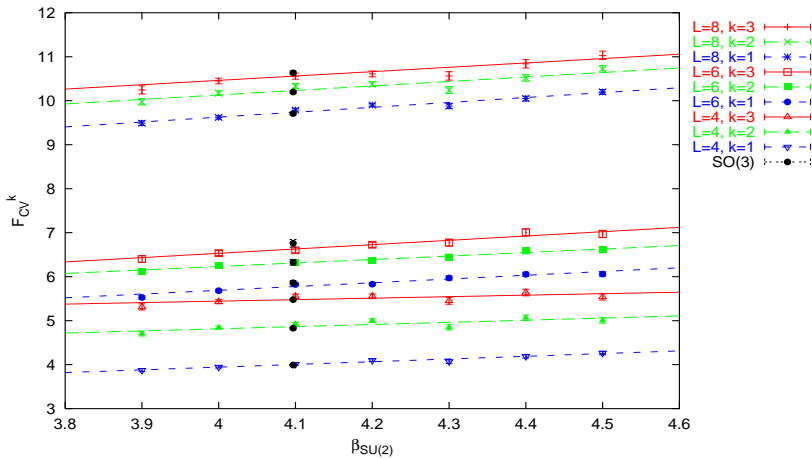}}
\caption{Comparison of the twist free energies between $SO(3)$ at $\beta=4.5$
and $SU(2)$ at various $\beta$'s. Electric twist in 1, 2 and 3 planes is considered,
on lattices of size $4^3, 6^3$ and $8^3 \times 4$. One finds
$\beta_{SO(3)}=4.5 \longleftrightarrow \beta_{SU(2)}=4.10(15)$.}
\label{match}
\end{figure}

The twisted free energies $F_{CV}^k$ are continuum quantities, which depend
upon the spatial size $L$ and the temperature $T$. If the $SO(3)$ and $SU(2)$
theories represent the same continuum physics, then to each $\beta_{SO(3)}$
should correspond a value $\beta_{SU(2)}$, which yields the same lattice
spacing and thereby the same $F_{CV}^k$ for equal lattice sizes, modulo
small lattice artifacts. We test this statement in Fig.~\ref{match}.
The 9 free energies $F_{CV}^k, k=1,2,3$ measured on $4^3, 6^3$ and $8^3 \times 4$
lattices in $SO(3)$ at $\beta_{SO(3)} = 4.5$ are compared with similar
quantities measured in $SU(2)$ at different values of $\beta_{SU(2)}$.
The straight lines show a linear interpolation in $\beta_{SU(2)}$ of the
$SU(2)$ data. One observes a very good match of all 9 observables,
for $\beta_{SU(2)} \approx 4.10(15)$.

Thus, $a^{-1}(\beta_{SO(3)}=4.5) \approx a^{-1}(\beta_{SU(2)}=4.1)$, which is
about $200$ GeV! As conventional wisdom asserts, our $SO(3)$ lattice is
very fine indeed. To reach low temperatures $T < T_c$ and probe the 
confined phase would require a lattice of size ${\cal O}(700)^4$, far
beyond what is currently achievable.

Note the closeness of the matched $SO(3)$--$SU(2)$ bare couplings
($\frac{4}{g^2} = 4.5$ vs $4.1$). This should come as no surprise.
Lattice perturbation theory is identical between the $SU(2)$ Wilson
action and the $SO(3)$ Villain action: the difference resides in the
$\alpha_P$-monopoles, which do not appear in the perturbative expansion.
Therefore, $\Lambda_{\rm{lattice}}$ is the same in both theories,
and one should expect similar values for the non-perturbatively matched
$\beta$'s. The $\alpha_P$-monopoles disorder the $SO(3)$ theory slightly,
which is why the matching $SU(2)$ $\beta$ is slightly smaller.

\section{Conclusion}

To summarize, conventional wisdom prevails and there is no mystery.
The $SO(3)$ lattice theory at weak coupling gives the same non-perturbative
physics as the $SU(2)$ theory and a common confinement order parameter
exists for both: the center vortex -- or twist -- free energy.
The difference is that a definite twist sector is selected via the choice
of boundary conditions in $SU(2)$, whereas all sectors are automatically
summed over in $SO(3)$.

Unfortunately, the numerical study of the $SO(3)$ theory seems limited to very 
fine lattice spacings, because a bulk transition prevents exploring 
continuum physics at stronger bare couplings. One way out is to suppress
the formation of monopoles whose condensation triggers the bulk transition
\cite{MMP}. However, this appealing approach suffers from an unpleasant,
technical side-effect. The Monte Carlo sampling over the various twist
sectors, which is essential to measure the confinement order parameter,
is made more difficult by higher action barriers separating the sectors.
This situation is analogous to simulations of overlap or domain-wall fermions.
There, dislocations separating topological sectors cause trouble in the
inversion of the Dirac operator. They can be suppressed by choosing a gauge
action which assigns them a large action. The price to pay is a much slower
Monte Carlo evolution of the topological charge. 
One may hope that a clever enhancement of the sampling probability in the
neighborhood of the saddle point, and only there, as in \cite{AvdS},
can solve both types of problems.

Finally, one can try to generalize the lesson learnt here about the
structure of the gauge group $G$ necessary for confinement. $SO(3)$
shows that a non-trivial center is not required. On the other hand,
the existence of the dual string tension arises from that of the twist
sectors. Those in turn follow from the non-trivial first homotopy
group $\Pi_1(G/Center(G))$, which is the same for $SU(2)$ and $SO(3)$.
A non-trivial $\Pi_1(G/Center(G))$, or more precisely $\Pi_1(G/Z_G)$,
where $Z_G$ is the discrete part of the center of $G$, is necessary
for a dual string tension to be defined. The non-trivial elements of
this group are the center vortex (or twist) excitations.  Therefore,
we conjecture that the existence of center vortices is a necessary
condition for the existence of an ordinary string tension, i.e.  for
confinement. It is not a sufficient condition: compact $U(1)$ has
twist sectors but its weak-coupling phase does not confine.  On the
other hand, there are non-Abelian Lie groups which do not meet this
condition, namely $G_2$, $F_4$ and $E_8$.  Therefore, we conjecture
that these cannot confine (in the sense that the Wilson loop cannot
obey an area law).

\section{Acknowledgements}
We are grateful to G.~Burgio, M.~M\"uller-Preussker and L.~von Smekal 
for discussions.
We thank the conference organizers for their talent in creating a stimulating
atmosphere and their patience in awaiting this manuscript.

\end{document}